\documentclass[%
 reprint,
superscriptaddress,
amsmath,amssymb,
aps,
]{revtex4-2}

\usepackage{graphicx}
\usepackage{dcolumn}
\usepackage{bm}
\usepackage{hyperref}
\usepackage{verbatim} 
\hypersetup{
colorlinks,
citecolor=blue,
filecolor=blue, 
linkcolor=blue, 
urlcolor=blue
}

\usepackage{color}
\usepackage{xcolor}
\usepackage{ulem}
\usepackage{physics}

\newcommand{\yambo}{\textsc{yambo}}
\newcommand{\qe}{\textsc{Quantum Espresso}}

\newcommand{\q}{{\mathbf q}}

\renewcommand{\Im}{\mathrm{Im}} 
\renewcommand{\Re}{\mathrm{Re}}

\newcommand{\editor}[2]{%
  \expandafter\newcommand\csname #1note\endcsname[1]{%
    \textcolor{#2}{(\textbf{#1:} \textit{##1})}}%
  \expandafter\newcommand\csname #1\endcsname[1]{%
    \textcolor{#2}{##1}}%
  \expandafter\newcommand\csname #1cancel\endcsname[1]{%
    \textcolor{#2}{\sout{##1}}}%
\expandafter\newcommand\csname #1can\endcsname[1]{%
    \textcolor{#2}{\sout{##1}}}%
  \expandafter\newcommand\csname #1change\endcsname[2]{%
    \textcolor{#2}{\sout{##1} ##2}}%
      \expandafter\newcommand\csname #1ch\endcsname[2]{%
    \textcolor{#2}{\sout{##1} ##2}}%
  \newenvironment{#1text}{\color{#2}}{\color{black}}
}
\definecolor{Blu}{rgb}{0.00,0.00,1.00}
\definecolor{Red}{rgb}{1.00,0.00,0.00}
\definecolor{Cyan}{rgb}{0.00,0.50,0.50}
\definecolor{Green}{rgb}{0.00,0.70,0.00}

\editor{DAL}{Green}
\editor{CC}{Cyan}
\editor{kb}{orange}
\editor{KB}{orange}
\editor{R}{Red}

\renewcommand{\emph}{\textit}
\newcommand{\suppinfo}{Ref.~\cite{supp-info}}

\begin{document}

\title{Plasmonic properties and correlation energies from a compact multipole representation of the dielectric response in 2D metals}

\author{Dario A. Leon}
\email{dario.alejandro.leon.valido@nmbu.no}
\affiliation{
 Department of Mechanical Engineering and Technology Management, \\ Norwegian University of Life Sciences, NO-1432 Ås, Norway
}%

\author{Claudia Cardoso}
\affiliation{
 S3 Centre, Istituto Nanoscienze, CNR, 41125 Modena, Italy
}

\author{Kristian Berland}
\affiliation{
 Department of Mechanical Engineering and Technology Management, \\ Norwegian University of Life Sciences, NO-1432 Ås, Norway
}

\date{\today}

\begin{abstract}
Multipole–Padé approximants provide a compact representation of dynamical response functions in terms of a small number of collective modes. Here, we generalize this framework to incorporate momentum dependence across the full Brillouin zone of 2D metals by constructing a symmetry-conserving, anisotropic representation of the inverse dielectric function. 
This analytic form enables efficient and accurate evaluation of quantities involving dynamical screening, including spectral features and correlation energies. 
We construct such compact representations for a set of seven two dimensional metals spanning distinct electronic regimes, and show that a small number of dispersive plasmonic modes suffices to accurately describe the dielectric response across the full Brillouin zone, while also yielding accurate correlation energies. 
The proposed representation therefore establishes a direct bridge between {\it ab initio} calculations and analytical models of screening, opening new avenues for applications in condensed matter systems. 
\end{abstract}

\maketitle

\section{Introduction}
%
The dielectric function describes the collective electronic response of a material and underlies a wide range of spectroscopic and many-body phenomena. Two-dimensional (2D) metals exhibit rich momentum-dependent screening and  plasmonic excitations due to reduced dimensionality and anisotropic electronic structure~\cite{Hwang2007PRB,Pfnur2020SIP,daJornada2020NatCom,Muniain2022PCCP,Cardoso2026PRB}. These systems have attracted broad interest due to their diverse potential applications~\cite{Jablan2013PIEEE,Chen2018ChemRev,Fan2019AOM,Wang2020MAT,Yousaf2021JPCC,Do2025NatComm}. 
First-principles calculations based on the Random Phase Approximation (RPA) provide access to the dielectric response over wide frequency ranges, allowing detailed comparison with experiments~\cite{martin2016book,Leon2026JPCM}. However, such a numerical approach typically yield large datasets that are difficult to interpret and impractical to use as a starting point for further modeling and analysis~\cite{Leon2026JPCM}.

Approaches based on multipole-Pad\'e approximants (MPA) provide a route toward compact, physically transparent, and accurate representations of dynamical response functions in terms of a small number of collective modes~\cite{Leon2021PRB,Leon2023PRB,Leon2025PRB,Leon2026PRB}. In particular, the recent momentum-dependent MPA($\q$) formulation~\cite{Leon2026PRB} has been shown to successfully capture complex plasmon dispersions for 26 distinct bulk elemental metals~\cite{Leon2026PRB}. However, MPA($\q$) has so far been applied only to one-dimensional momentum paths, which, while useful for interpretation, is not suited for evaluating properties requiring integration over the full Brillouin zone. 

In this work, we extend the MPA($\q$) framework to 
the full 2D Brillouin zone by constructing symmetry-conserving multipole–Padé representations of the inverse dielectric function, $\varepsilon^{-1}(\q, \omega)$, over a wide energy range.
The residues and plasmonic frequencies of the resulting analytical models are expressed as smooth functions of momentum $\q$, defining band-like structures in the spectral representation of the dielectric response. 
We apply the method to a representative set of 2D metals spanning distinct electronic regimes, including nearly free-electron systems such as Na, K, and Mg; layered borides such as MgB$_2$ and AlB$_2$ with multiband $\sigma$ and $\pi$ carriers~\cite{Yousaf2021JPCC}; the transition metal dichalcogenide NbSe$_2$~\cite{Wang2017NatComm}; and the electride Ca$_2$N~\cite{Guan2015SciRep,Cudazzo2017PRB}, characterized by loosely bound interlayer electrons. Despite the diversity of these systems, we find that only a small number of dispersive plasmonic modes suffice to accurately capture the main features of the full momentum and frequency dependent dielectric response. 

We demonstrate that the RPA correlation energy~\cite{Eshuis2012TCA,Ren2012JMS} can be efficiently computed using the compact MPA($\q$) representation evaluated on the imaginary frequency axis. 
In contrast to direct first-principles approaches based on discrete frequency and momentum grids, the analytical MPA($\q$) model interpolates between the grid points, improving the description of the $\q \to 0$ limit, and enabling efficient evaluation on dense momentum samplings that would otherwise be computationally prohibitive. By replacing large numerical datasets with compact analytical forms, the resulting MPA($\q$) models establish a direct bridge between large-scale {\it ab initio} calculations and  analytical modeling of dynamical screening. More broadly, MPA($\q$) can facilitate efficient implementations of many-body approaches requiring momentum- and frequency-dependent dielectric functions, including GW/BSE calculations and coupled quasiparticle frameworks~\cite{Mahan2013book,martin2016book}.

\section{Methods}
%
\subsection{RPA dielectric response and correlation energy}
Within the RPA approximation, the interacting polarizability $\chi$ is obtained from the Dyson equation
\begin{equation}
    \chi = \chi_0 + \chi_0 v \chi,
\end{equation}
where $v$ is the Coulomb interaction and $\chi_0$ is the independent-particle polarizability. The inverse dielectric matrix is then given by
\begin{equation}
    \varepsilon^{-1} = 1 + v \chi.
    \label{eq:dyson}
\end{equation}

The RPA correlation energy~\cite{Eshuis2012TCA,Ren2012JMS} is evaluated as
\begin{equation}
    E_c = \frac{\hbar}{2\pi}\int \frac{d \q^2}{(2 \pi)^2} \int_0^{\infty} d\omega [\ln{\varepsilon(\q,i \omega)} -\varepsilon(\q,i \omega) + 1],
    \label{eq:Ec}
\end{equation}
where the macroscopic dielectric function is defined as
\begin{equation}
\varepsilon(\q,\omega)
=
[\epsilon^{-1}_{00}(\q,\omega)]^{-1},
\end{equation}
with $\epsilon^{-1}_{00}$ denoting the head of the inverse dielectric matrix. Local-field effects are therefore included through the inversion of the full dielectric matrix~\cite{Leon2026JPCM}.

\subsection{The MPA($\q$) model}
The MPA approach~\cite{Leon2021PRB,Leon2023PRB,Leon2025PRB} provides an efficient representation of the frequency dependence of the screened interaction by expressing it as a sum over a finite number of poles. 
In contrast to single plasmon-pole models, MPA systematically improves accuracy with an increasing number of poles, rapidly reaching full-frequency accuracy. 
The more recent MPA($\q$)~\cite{Leon2026PRB} explicitly incorporates the momentum dependence as 
\begin{equation}
    \varepsilon^{-1}_{\mathrm{MPA}}(\q, \omega) = 1 + \sum_p^{n_{p}} \frac{2 R_p (\q) \Omega_p(\q)}{\omega^2 - [\Omega_p(\q)]^2},
    \label{eq:mpa_model}
\end{equation}
where $\q$ lies in the first BZ, and $R_p(\q)$ and $\Omega_p(\q)$ are complex, smooth functions of $\q$. The pole positions satisfy the causality condition $-\Re[\Omega_p(\q)] < \Im[\Omega_p(\q)] < 0$.

In Ref.~\cite{Leon2026PRB} we fitted the MPA($\q$) model along selected high-symmetry $\q$-paths, using third-order polynomials in $q \equiv |\q|$ for $R_p(q)$ and $\Omega_p(q)$. Here, we generalize this framework to the full 2D Brillouin zone by constructing a symmetry-conserving model accounting for isotropic and anisotropic terms in polar coordinates ($q, \theta$):
\begin{equation}
\begin{split}
    R_p(\q) = R_p [f_{R, p}^\mathrm{iso}(q) + f_{R, p}^\mathrm{ani}(q, \theta)] \\
    \Omega_p(\q) = \Omega_p [f_{\Omega, p}^\mathrm{iso}(q) + f_{\Omega, p}^\mathrm{ani}(q, \theta)].
\end{split}
\end{equation}

For the isotropic terms, here we also adopt a third-order polynomial:
\begin{equation}
\begin{split}
    f_{R, p}^\mathrm{iso}(q) = 1 + a_{1, p} q + a_{2, p} q^2 + a_{3, p} q^3 \\
    f_{\Omega, p}^\mathrm{iso}(q) = 1 + b_{1, p} q + b_{2, p} q^2 + b_{3, p} q^3,
\end{split}
\label{eq:ERiso}
\end{equation}
which captures the smooth momentum dependence observed in first-principles data. This form yields a finite plasmon energy at $q \to 0$, due to interband contributions 
to the dielectric response in real materials. Note that for intraband contributions, the vanishing $\sqrt{q}$ dispersion expected from an ideal 2D homogeneous electron gas arises at very small momenta~\cite{daJornada2020NatCom}. Although this regime could be incorporated through an appropriate modification of the low-$q$ parametrization, we neglect it here, since it is relevant only for momenta below the resolution of the present calculations. 

For the anisotropic terms, we use a simple symmetry-conserving rational form: 
\begin{equation}
\begin{split}
    f_{R, p}^\mathrm{ani}(q, \theta) = \cos(n \theta) \frac{A_p q^2}{1 + B_p q} \\
    f_{\Omega, p}^\mathrm{ani}(q, \theta) = \cos(n \theta) \frac{C_p q^2}{1 + D_p q},
\end{split}
\label{eq:ERani}
\end{equation}
which preserves analyticity at $q \to 0$ and captures the leading angular dependence. The integer $n$ is fixed by the rotational symmetry of the lattice (e.g., $n=4$ and $n=6$ for tetragonal and hexagonal systems, respectively). 
Higher-order harmonic terms were not required to describe the dielectric response within the aimed accuracy. 

All coefficients $a_{i, p}$, $b_{i, p}$, $A_p$, $B_p$, $C_p$, and $D_p$ are obtained by fitting the calculated first-principles inverse dielectric function, resulting in a total of $12 n_p$ parameters.

\subsection{Computational details}
\label{sec:com}
Density functional theory (DFT) calculations were performed using the {\qe}~\cite{QE1, QE2} plane wave package with the Perdew-Burke-Ernzerhof variant of the generalized gradient approximation functional~\cite{Perdew1996PRL}. 
We adopted the norm-conserving optimized Vanderbilt pseudopotentials of Ref.~\cite{Hamann2013PRB}, with a kinetic plane wave energy cutoff of 70 Ry. 
The Brillouin zone was sampled with a $24\times 24\times 1$ $\Gamma$-centered Monkhorst-Pack grid. Metallic occupations were treated using ordinary Gaussian smearing with a width of 0.01 Ry. 
Calculations of dielectric spectra within the RPA approximation on top of DFT were performed with the {\yambo} code~\cite{Marini2009CPC,Sangalli2019JPCM}. The RPA response was evaluated in the time-ordered formulation with a kinetic plane wave energy cutoff of 10 Ry and a total of 100 bands. An homogeneous grid of 500 points in an interval of 100 eV was used to evaluate the response along both the real and imaginary frequency axes. A damping of $0.2$~eV was used when sampling the response along the real frequency axis.

\section{Results}
%
The set of 2D metals studied here and their lattice properties are listed in Table~\ref{tab:2Dmetals}.
\begin{table}[hbt!]
  \begin{ruledtabular}
    \begin{tabular}{cccccc} 
      \\[-8pt]
      2D metal & space group & a (\AA) & $n_p$ & $\Delta E_c$ (meV) & $E_c^{\mathrm{ext}}$ (meV)\\
      \hline\\[-8pt]     
            Na & P6/mmm  & 3.5484 & 1 & 0.96 & -67.6 \\
            Mg & P4/mmm  & 3.0558 & 2 & 0.09 & -292.2 \\
             K & P6/mmm  & 4.5760 & 2 & 0.49 & -88.7 \\
       MgB$_2$ & P6/mmm  & 3.0634 & 4 & 0.77 & -110.5 \\
       AlB$_2$ & P6/mmm  & 2.9989 & 4 & 1.29 & -146.1 \\
      NbSe$_2$ & P6/mmm  & 3.4660 & 5 & 0.07 & -452.1 \\
       Ca$_2$N & P6/mmm  & 3.6127 & 6 & 1.22 & -418.9 \\
    \end{tabular}
  \end{ruledtabular}
\caption{2D metals studied in this work. Symmetry groups and lattice parameters, $a$, were obtained from Refs.~\cite{Mounet2018NatNano,Campi2023ACSnano}. Here, $n_p$ is the number of poles used to fit the MPA($\q$) model, $\Delta E_c$ is the deviation of the correlation energy computed with MPA($\q$) with respect to a numerical evaluation for the $24\times 24\times 1$ $\q$-grid, and $E_c^{\mathrm{ext}}$ the extrapolated value of the model for infinitely dense grids. All the parameters of the model for each material are reported in \suppinfo.
}
    \label{tab:2Dmetals}
\end{table}

\subsection{Spectral representation}
\begin{figure*}[hbt!]
    \centering
    \includegraphics[width=0.5\textwidth]{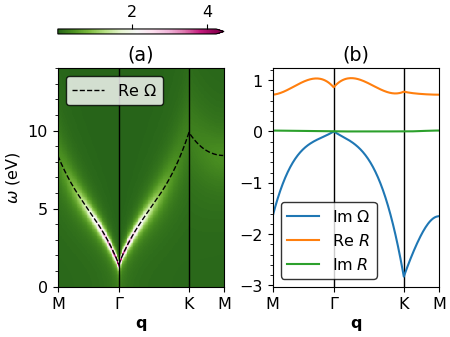}
        \includegraphics[width=0.235\textwidth]{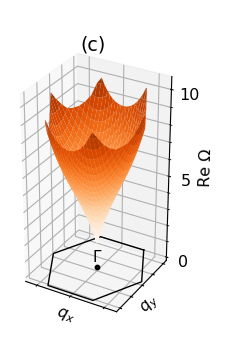}
        \includegraphics[width=0.235\textwidth]{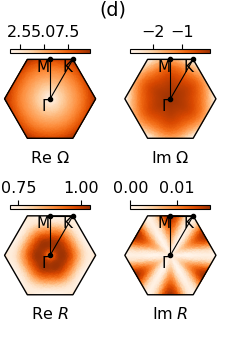}
    \caption{Spectral properties of the dielectric response of monolayer Na obtained with MPA($\q$) with one pole, $n_p=1$. (a) SBS corresponding to $|\Im [\epsilon^{-1} (\omega, \q)]|$ along the $M \Gamma X M$ high-symmetry path, with dashed lines indicating
    the plasmon dispersion $\Re [\Omega_p (\q)]$.
    (b) Bands corresponding to $\Im [\Omega_p (\q)]$, $\Re [R_p (\q)]$, and $\Im [R_p (\q)]$ along the same high-symmetry path. (c) Surface plot of the plasmon energy $\Re [\Omega_p (\q)]$ as a function of $\q$ in the full hexagonal 2D Brillouin zone. Colormaps in (d) show the degree of anisotropy of $\Omega_p (\q)$ and $R_p (\q)$ and their symmetry. The scale of intensities is relative to each plot, going from minimum (light tones) to maximum (dark tones).}
    \label{fig1}
\end{figure*}

We start by applying MPA($\q$) to monolayer Na as a prototypical system well described with a single plasmon pole. Fig.~\ref{fig1}(a) shows the $|\Im[\epsilon^{-1}(\omega,\q)]|$ spectral band structure (SBS) obtained with MPA($\q$) along a high-symmetry $\q$-path. The maximum intensity follows the dispersion of the plasmon given by $\Re[\Omega_p(\q)]$. Fig.~\ref{fig1}(b) shows analogous bands for $\Im[\Omega_p(\q)]$, $\Re[R_p(\q)]$ and $\Im[R_p(\q)]$, along the same $\q$-path. The full $\q$ dependence of the plasmon energy is shown in Fig.~\ref{fig1}(c), where $\Re[\Omega_p(\q)]$ defines a continuous plasmonic band surface over the 2D Brillouin zone. The corresponding maps of $\Omega_p(\q)$ and $R_p(\q)$ in Fig.~\ref{fig1}(d) reveal the expected symmetry and quantify the degree of anisotropy, which remains weak for this nearly isotropic system.

\begin{figure*}[hbt!]
    \centering
    \includegraphics[width=0.99\textwidth]{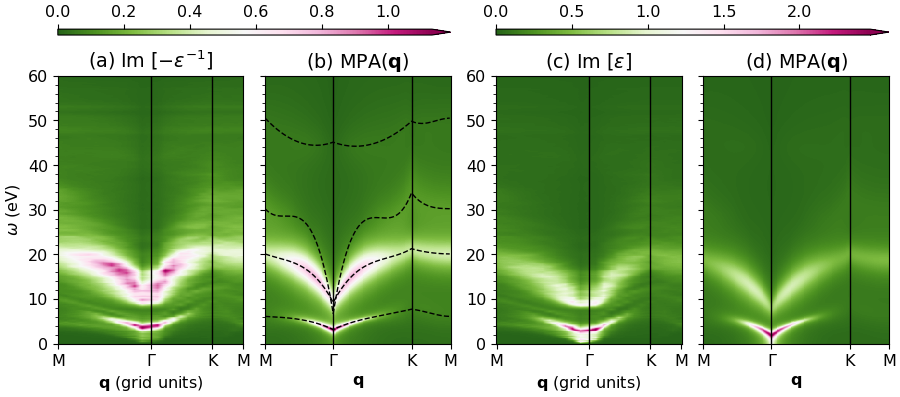}
    \caption{Comparison of the $|\Im [\epsilon^{-1} (\omega, \q)]|$ (a, b) and $|\Im [\epsilon (\omega, \q)]|$ (c, d) SBSs of monolayer MgB$_2$, constructed from the numerical RPA data (a, c) and its corresponding MPA($\q$) representation with $n_p=4$ poles (b, d). For visualization purposes, the numerical data are plotted in grid units, with segment lengths proportional to the number of sampled points along the high-symmetry path of the $24\times24\times1$ $\q$-grid. In addition, the finite sampling limits the description of the $\q \to 0$ region in the numerical data, while the MPA($\q$) model provides a continuous representation.}
    \label{fig2}
\end{figure*}

Next, we assess the performance of MPA($\q$) for more complex systems. As an illustrative example, Fig.~\ref{fig2} compares both $|\Im [\epsilon^{-1} (\omega, \q)]|$ (a, b) and $|\Im [\epsilon (\omega, \q)]|$ (c, d) SBSs of monolayer MgB$_2$ obtained from first-principles calculations (a, c) with the fitted MPA($\q$) model with $n_p=4$ poles (b, d). The model reproduces the main features in both the direct and inverse dielectric functions throughout the entire energy range. Higher accuracy can be systematically achieved by increasing $n_p$.

The number of poles used for each material is summarized in Table~\ref{tab:2Dmetals}. Across all systems considered, we find that a small $n_p$ (typically between one and six) suffices to reproduce the full dielectric response, capturing fundamental spectral properties such as plasmonic dispersions and damping. Simple metals such as Na, Mg, and K are accurately described by one or two poles, while more complex materials such as MgB$_2$, AlB$_2$, NbSe$_2$, and Ca$_2$N require additional poles to capture multiple plasmonic branches.
Moreover, the symmetry-conserving angular dependence captures anisotropic dispersion while preserving the underlying lattice symmetry, providing a compact and physically transparent description of collective excitations in 2D metals.

\subsection{Correlation energy}
\begin{figure}
    \centering
    \includegraphics[width=0.5\textwidth]{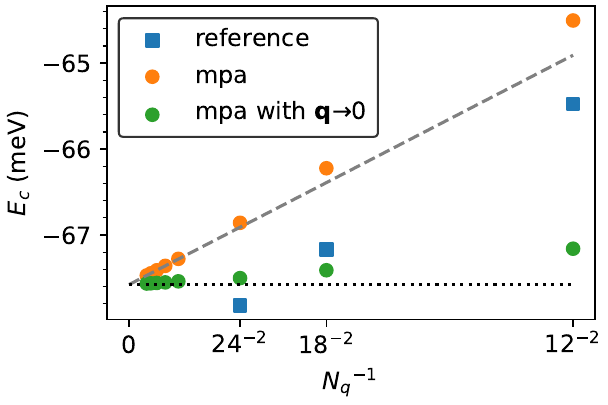}
    \caption{Convergence of the RPA correlation energy $E_c$ of monolayer Na as a function of the number of sampling points, $N_q$, of the Monkhorst-Pack $\q$-grid. MPA($\q$) (yellow circles) reproduces the reference {\it ab initio} data (blue squares) with an uniform error of $\sim 1$ meV. The MPA($\q$) representation enables the evaluation of $E_c$ on denser grids, fitted with a linear dashed line, which allows for accurate extrapolation to the dense-grid limit (dotted line). The improved $\q \to 0$ contribution within MPA($\q$) (green circles) significantly accelerates convergence toward the extrapolated value. 
    }
    \label{fig3}
\end{figure}

We next assess the accuracy of the MPA($\q$) representation for evaluating the RPA correlation energy of Eq.~\eqref{eq:Ec}. We compute $E_c$ using both the numerical first-principles dielectric function and its corresponding MPA($\q$) representation evaluated on the same $\q$-grid. In Table~\ref{tab:2Dmetals} we report the resulting difference for all systems studied, which is always below $1.3$ meV, demonstrating that a small number of poles is sufficient to accurately reproduce the correlation energy of all systems studied.

A key advantage of having a smooth analytical representation of $\varepsilon(\q,i\omega)$ is that 
Eq.~\eqref{eq:Ec} can be evaluated on flexible $\q$ and $\omega$ grids. In addition, the parametrization obtained from finite $q$ elements improves the description of the $\q \to 0$ limit, by including  contributions that are not straightforwardly captured in standard {\it ab initio} calculations, due to the sharpness of the dielectric response of low-dimensional materials~\cite{Guandalini2024PRB,Sesti2025Arxiv}. 
In particular, the analytical momentum dependence enables systematic refinement of the Brillouin-zone integration beyond the original {\it ab initio} grid, and extrapolation toward the dense-grid limit.

The corresponding extrapolated values of $E_c$, obtained from MPA($\q$) with denser $\q$-grids, are reported in Table~\ref{tab:2Dmetals}. Figure~\ref{fig3} illustrates this procedure for monolayer Na, showing that even if fitted on a fixed $\q$-grid, MPA($\q$) reproduces the convergence trend. While the difference between the extrapolated value and the value obtained on the original $24\times24\times1$ $\q$-grid is small for Na, this is not always the case. Moreover, the improved description of the $\q \to 0$ limit significantly accelerates the convergence for all the systems studied. 
These results show that MPA($\q$) is not only a compact representation of the dielectric response, but also a practical framework for efficiently evaluating quantities involving dynamical screening.

\section{Conclusions}
We have introduced a symmetry-conserving extension of the MPA($\q$) framework to the full Brillouin zone, and demonstrated its accuracy and efficiency for a representative set of 2D metals. The resulting MPA($\q$) representation provides a compact description of the dielectric response, capturing both spectral features and correlation energies with a small number of plasmonic poles. 
The smooth analytical dependence across the entire Brillouin zone enables controlled and systematic refinement of the momentum sampling and the $\q \to 0$ limit, essential in metallic systems and low-dimensional materials where the dielectric response exhibits strong nonlocal behavior. As a result, momentum-space integrals such as the RPA correlation energy can be evaluated with significantly improved convergence, as demonstrated in this work. 
Our results establish a direct bridge between large-scale {\it ab initio} calculations and analytical models retaining predictive accuracy and enabling efficient reuse of the data. 
This provides a practical route for efficiently evaluating screening observables within RPA and offers a foundation for extensions to many-body approaches beyond. 

\section*{Data Availability}
The data generated in this article are openly available on Ref.~\cite{MPAq_data}.

\section*{Author Contributions}
D.A.L. conceived the idea of the work, developed and implemented the methods, and wrote the first draft. 
C.C. performed the first-principle calculations. All authors contributed to the analysis of the results, discussion, and manuscript writing. 

%
%
\bibliography{biblio}

\end{document}